\documentclass{emulateapj}

\usepackage{rotate}
\usepackage{longtable}
\usepackage{color}

\shorttitle{Host Morphologies of Compton-Thick AGN at $z\sim1$}
\shortauthors{Kocevski et al.}

\begin{document}

\title{Are Compton-Thick AGN the Missing Link Between Mergers and Black Hole Growth?}

\author{Dale D.~Kocevski\altaffilmark{1}, 
Murray Brightman\altaffilmark{2}, 
Kirpal Nandra\altaffilmark{3}, 
Anton M.~Koekemoer\altaffilmark{4}, 
Mara Salvato\altaffilmark{3}, 
James Aird\altaffilmark{5}, 
Eric F.~Bell\altaffilmark{6}, 
Li-Ting Hsu\altaffilmark{3}, 
Jeyhan S.~Kartaltepe\altaffilmark{7}, 
David C.~Koo\altaffilmark{8}, 
Jennifer M.~Lotz\altaffilmark{4}, 
Daniel H.~McIntosh\altaffilmark{9}, 
Mark Mozena\altaffilmark{8}, 
David Rosario\altaffilmark{3}, 
Jonathan R.~Trump\altaffilmark{10}}

\affil{Department of Physics and Astronomy, Colby College, Waterville, ME 04961}
\altaffiltext{2}{California Institute of Technology}
\altaffiltext{3}{Max-Planck-Institut f\"ur Extraterrestrische Physik}
\altaffiltext{4}{Space Telescope Science Institute}
\altaffiltext{5}{Institute of Astronomy, Cambridge}
\altaffiltext{6}{University of Michigan}
\altaffiltext{7}{National Optical Astronomy Observatories}
\altaffiltext{8}{University of California, Santa Cruz}
\altaffiltext{9}{University of Missouri, Kansas City}
\altaffiltext{10}{The Pennsylvania State University}

\email{dale.kocevski@colby.edu}

\begin{abstract}

We examine the host morphologies of heavily obscured active galactic nuclei (AGN) at $z\sim1$ to test whether obscured supermassive black hole growth at this epoch is preferentially linked to galaxy mergers.  Our sample consists of 154 obscured AGN with $N_{\rm H}>10^{23.5}$ cm$^{-2}$ and $z<1.5$.  Using visual classifications, we compare the morphologies of these AGN to control samples of moderately obscured ($10^{22}$ cm$^{-2}$ $<N_{\rm H}< 10^{23.5}$ cm$^{-2}$) and unobscured ($N_{\rm H}<10^{22}$ cm$^{-2}$) AGN.  These control AGN have similar redshifts and intrinsic X-ray luminosities to our heavily obscured AGN.  We find that heavily obscured AGN are twice as likely to be hosted by late-type galaxies relative to unobscured AGN ($65.3^{+4.1}_{-4.6}\%$ vs $34.5^{+2.9}_{-2.7}\%$) and three times as likely to exhibit merger or interaction signatures ($21.5^{+4.2}_{-3.3}\%$  vs $7.8^{+1.9}_{-1.3}\%$).  The increased merger fraction is significant at the 3.8$\sigma$ level.  If we exclude all point sources and consider only extended hosts, we find the correlation between merger fraction and obscuration is still evident, however at a reduced statistical significance ($2.5\sigma$).  The fact that we observe a different disk/spheroid fraction versus obscuration indicates that viewing angle cannot be the only thing differentiating our three AGN samples, as a simple unification model would suggest.  The increased fraction of disturbed morphologies with obscuration supports an evolutionary scenario, in which Compton-thick AGN are a distinct phase of obscured SMBH growth following a merger/interaction event.  Our findings also suggest that some of the merger-triggered SMBH growth predicted by recent AGN fueling models may be hidden among the heavily obscured, Compton-thick population.

\end{abstract}

\keywords{galaxies: active --- galaxies: evolution --- X-rays: galaxies}

\section{Introduction}

Observations over the past two decades have revealed a tight correlation between the mass of a galaxy's stellar bulge and its central super-massive black hole (SMBH; Magorrian et al. 1998; Gebhardt et al.~2000; Tremaine et al.~2002, G{\"u}ltekin et al.~2009; McConnell \& Ma 2013).  
This finding is commonly interpreted as evidence that the growth of SMBHs and their host spheroids is connected.  
Given the effectiveness of violent galaxy mergers to dissipate angular momentum, mergers have long been proposed as a means to forge this connection (Sanders et al.~1988; Hernquist et al.~1989; Kauffmann \& Haehnelt 2000).  In this scenario, the strong gravitational torques produced as a result of a merger funnel gas to the center of a galaxy, triggering both accretion onto the central black hole and star formation that grows the stellar bulge (Barnes \& Hernquist 1991; Mihos \& Hernquist 1996).  Coupled with self-regulated black hole growth (i.e.,~AGN feedback; Di Matteo et al.~2005; Hopkins et al.~2005, 2006), galaxy mergers provide an attractive mechanism to both trigger AGN activity and help explain the co-evolution observed between SMBHs and their host galaxies.


However, observational attempts to tie AGN activity to galaxy mergers have produced mixed results.  Gas-rich mergers are observed to fuel a substantial fraction of bright quasars (e.g.~Guyon et al.~2006; Bennert et al.~2008; Veilleux et al.~2009; Koss et al.~2010, 2012) and recent studies of kinematic galaxy pairs have demonstrated that nuclear activity is indeed enhanced in galaxies with an interacting companion (Silverman et al.~2011; Ellison et al.~2011).  On the other hand, morphological studies have consistently found that the bulk of the AGN population does not appear to be triggered by major galaxy mergers.  Both at $z\sim1$ (Grogin et al.~2005; Pierce et al.~2007; Cisternas et al.~2011; Villforth et al.~2014) and more recently at $z\sim2$ (Schawinski et al.~2011; Kocevski et al.~2012; Rosario et al.~2015), studies have found that X-ray selected AGN hosts are no more likely to exhibit morphological disturbances compared to similar inactive galaxies.  In fact, results from the CANDELS survey (Grogin et al.~2011; Koekemoer et al.~2011) indicate that roughly half of moderate-luminosity ($L_{\rm X}<10^{43}$ erg s$^{-1}$) AGN at $z\sim2$ reside in disks and are likely fueled stochastically by secular processes and/or disk instabilities rather than major mergers (Kocevski et al.~2012).  


While the efficiency of stochastic fueling is expected to increase with redshift, given the rapid rise in the gas fraction of galaxies at $z>1$ (see e.g.~Tacconi et al.~2010), AGN fueling models predict that only a small fraction ($\sim$30\%) of the overall AGN luminosity density and BH mass density are the result of this fueling mode (Hopkins, Kocevski, \& Bundy 2014).  Instead the majority of SMBH growth is predicted to be the result of merger-induced fueling, especially at high luminosities ($L_{\rm X}>10^{44}$ erg s$^{-1}$; Hopkins \& Hernquist 2009; Draper \& Balantyne 2012).  The low merger fraction observed among X-ray selected AGN out to $z\sim2$ appears to be at odds with this prediction.

A major caveat associated with these findings is that heavily obscured AGN are not well sampled by X-ray surveys (see e.g.~Treister et al.~2004).  The most obscured, Compton-thick AGN (hereafter CT-AGN) are hidden by extreme column densities ($N_{\rm H} > 10^{24}$ cm$^{-2}$) of obscuring gas that can absorb even hard X-ray photons.  Analysis of the diffuse X-ray background indicates a significant fraction (up to $\sim50$\%) of AGN are hidden behind Compton-thick obscuration (Comastri et al.~1995; Ueda et al.~2003; Gilli et al.~2007); Akylas et al.~2012)\footnotemark[1], however much remains unknown about the demographics of their host galaxies.  In the evolutionary sequence of Sanders et al.~(1998), heavily obscured AGN represent a key phase in the life cycle of galaxies, as it is during this period that SMBHs are predicted to accrete the bulk of their mass (e.g., Fabian 1999; Gilli et al.~2007; Treister et al.~2009; Draper \& Ballantyne 2010).  Furthermore, hydrodynamical merger simulations predict that this obscured phase should coincide with the most morphologically disturbed phase of a galaxy interaction (Cattaneo et al.~2005; Hopkins et al.~2008). It is therefore acutely possible that past studies have systematically missed the AGN-merger connection by not sampling the obscured AGN population well.
\footnotetext[1]{Studies of resolved X-ray sources estimate a Compton-thick fraction of 35-40\% at $z>1$ (Brightman \& Ueda 2012; Buchner et al.~2015)}

Several studies have attempted to overcome this bias by selecting AGN at mid-infrared (IR) wavelengths, where radiation absorbed by obscuring circumnuclear dust is expected to be re-emitted (e.g.~Lacy et al.~2004; Stern et al.~2005; Daddi et al.~2007; Donley et al.~2007; Soifer et al.~2008).  
However, the most recent work to examine the morphologies of IR-selected AGN have produced conflicting results.
Schawinski et al.~(2012) examined the morphologies of $24\mu$m-selected Dust Obscured Galaxies (DOGs) at 
$z\sim2$, a high fraction of which are thought to host heavily obscured AGN based on X-ray stacking analyses (Fiore et al.~2008; Treister et al.~2009). The authors report a high disk fraction (90\%) and a relatively low merger fraction (4\%) that is consistent with studies of more unobscured AGN hosts (i..e.,~Schawinski et al.~2011).  On the other hand, Donley et al.~(2015) find that galaxies with a power-law spectral slope in the mid-IR, a signature of hot dust near an obscured AGN's central engine (Donley et al.~2007, 2012), have a higher fraction of disturbed morphologies compared to X-ray detected AGN hosts that do not exhibit similar IR emission.  This might be a result of the power-law technique preferentially selecting relatively high luminosity AGN, which may be more associated with galaxy mergers (e.g., Draper \& Ballantyne 2012; Treister et al. 2012).




In this study, we re-examine the connection between AGN obscuration and host morphology using a sample of heavily obscured AGN identified by their X-ray spectral properties.  Due to the differential absorption of hard and soft X-ray photons, the shape of an AGN's X-ray spectrum reveals not only the presence of gas along the line of sight, but it also provides a measure of its column density.  CT-AGN, in particular, can be identified by their X-ray spectra due to nuclear emission that is Compton scattered into our line of sight even when the direct emission is suppressed.  This ``reflected'' emission has a characteristic spectral shape consisting of a flat continuum and a high equivalent width Fe K$\alpha$ fluorescence line (Reynolds et al.~1994; Matt, Brandt \& Fabian 1996). 

Identifying CT-AGN using low energy ($<10$ keV) observations from \emph{Chandra} or \emph{XMM-Newton} is challenging because the heavy attenuation suffered at these wavelengths often restricts the accuracy of any X-ray spectral analysis.  In addition, CT-AGN often appear softer than expected at low energies due to their reflection-dominated emission.  As a result, a simple absorbed power-law fit to the soft X-ray spectra of CT-AGN will systematically underestimate their obscuring column density, as was recently shown using \emph{NuSTAR} observations  (Gandhi et al.~2014; Lansbury et al.~2015).  However, with sufficiently deep observations and proper spectral modeling, even the most obscured CT-AGN can be identified with relatively soft X-ray observations (e.g.~Brightman et al.~2014; Buchner et al.~2014).  Several studies have successfully employed X-ray spectral modeling to identify heavily obscured AGN using both deep \emph{Chandra} (Tozzi et al.~2006; Georgantopoulos et al.~2009; Feruglio et al.~2011; Gilli et al.~2011; Alexander et al.~2011; Brightman et al.~2014; Buchner et al.~2014) and \emph{XMM-Newton} observations (Comastri et al.~2011; Georgantopoulos et al.~2013; Lanzuisi et al.~2015).

\begin{figure}[t]
\epsscale{1.15}
\plotone{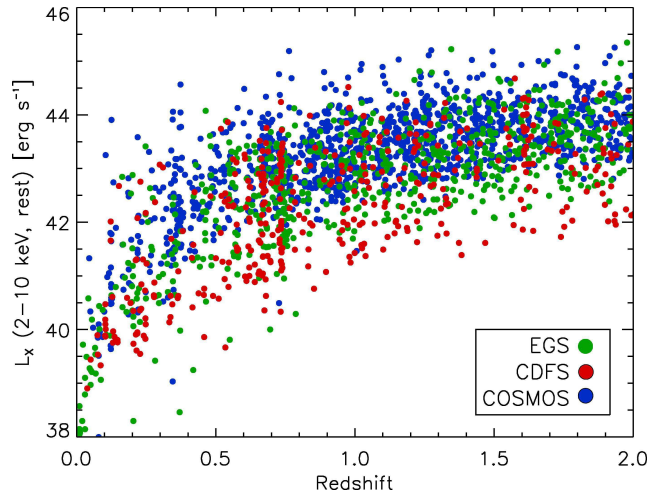}
\caption{Redshift versus luminosity for X-ray sources detected in our three target fields.  X-ray luminosities are intrinsic rest-frame 2-10 keV luminosities having been corrected for absorption (i.e.~setting $N_{\rm H}$=0 in our best-fit spectral model).  The use of both deep and wide survey data allows us to probe a wide range of luminosities, including sources at $L_{\rm X}>10^{44}$ erg s$^{-1}$, which are not well sampled in deep/narrow surveys such as the CDFS 4Ms observations. \label{fig-lx_z}}  
\end{figure}

For this work, we examine the host morphologies of the CT-AGN sample of Brightman et al.~(2014).  These sources were identified using the new spectral models of Brightman \& Nandra (2011) that correctly account for emission from Compton scattering, the geometry of the absorbing material, and include a self-consistent treatment for Fe K$\alpha$ emission. These models include all of the signatures of Compton-thick obscuration in a single model, allowing for the identification of CT-AGN in lower signal-to-noise data than previously possible.  Using visual classifications, we examine whether heavily obscured AGN exhibit an enhancement of merger and/or interaction signatures relative to their unobscured counterparts with the same intrinsic X-ray luminosity and redshift.  

Our analysis is presented as follows.  In \S2 we describe the X-ray and optical data used for the study, as well as discuss the methodology employed to select our sample of obscured AGN and unobscured control AGN.  The details of our morphological classification scheme are given in \S3 and our primary results are presented in \S4.  We discuss the implications of our findings in \S5.  Finally, our conclusions are summarized in \S6.  When necessary, the following cosmological parameters are used: $H_{0} = 70 {\rm kms^{-1}
  Mpc^{-1}; \Omega_{tot}, \Omega_{\Lambda}, \Omega_{m} = 1, 0.3, 0.7}$.


\section{Observations and Sample Selection}

\subsection{X-ray Datasets}

\begin{figure*}[t]
\epsscale{1.15}
\plotone{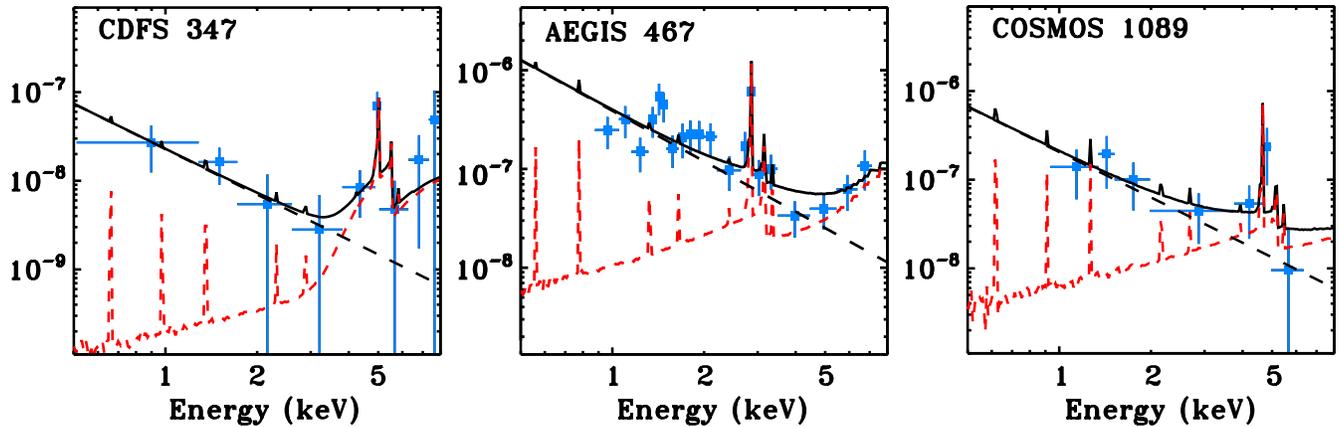}
\caption{X-ray spectra of three Compton-thick AGN detected in the CDFS, EGS, and COSMOS fields by Brightman et al.~(2014).  The red dashed line shows the best-fit direct torus emission from the AGN, while the black dashed line shows the Thompson scattered component. Due to heavy obscuration, the scattered component often dominates the emission at low energies, despite typically accounting for $<1$\% of the direct emission from these sources. All three sources exhibit strong Fe K$\alpha$ emission characteristic of a Compton-thick AGN.\label{fig-spectra}}  
\end{figure*}

The AGN sample used for our analysis is drawn from \emph{Chandra} datasets in three fields: the \emph{Chandra} Deep Field South (CDFS; Alexander et al. 2003; Xue et al.~2011), the AEGIS-XD dataset in the Extended Groth Strip (EGS; Nandra et al.~2015), and the C-COSMOS observations (Elvis et al.~2009; Civano et al.~2011).  These datasets have characteristic exposure times of 4 Msec, 800 ksec, and 180 ksec and cover an area of roughly 0.13, 0.28, 0.98 degrees$^{2}$, respectively.  This combination of deep and wide survey data was chosen to ensure that both moderate and high luminosity ($L_{\rm X}\sim10^{43-45}$ erg/s) AGN are well represented in our final sample.

X-ray source catalogs in the CDFS and AEGIS-XD were created by processing the \emph{Chandra} observations in each field with the custom reduction and source detection pipeline of Laird et al.~(2009).  These catalogs were matched to optical counterparts using the maximum-likelihood technique described by Sutherland \& Saunders (1992).  In the CDFS, the X-ray sources were cross-matched to the CANDELS F160W-selected photometry catalog of Guo et al.~(2013), while the AEGIS-XD sources were matched to the 3.6$\mu$m selected multi-waveband photometric catalog provided by the Rainbow Cosmological Surveys Database (Barro et al.~2011a,b).  For the C-COSMOS dataset, we adopt the published X-ray source and counterpart catalog of Civano et al.~(2012).  These published counterparts were identified by cross-matching the X-ray source catalog to the \emph{I}-band optical sample of Capak et al.~(2007) and the 3.6$\mu$m sample of Sanders et al.~(2007).


Redshifts for the identified X-ray counterparts were drawn from various spectroscopic datasets in each field.  For the CDFS, we used the compilation of Cardamone et al.~(2010) and Xue et al.~(2011).  For the EGS field, spectroscopic redshifts are drawn primarily from the DEEP2 (Newman et al.~2013) and DEEP3 (Cooper et al.~2012) redshift surveys.  For sources without spectroscopic redshifts in these fields, we use photometric redshifts from Hsu et al.~(2014) and Nandra et al.~(2015), which are derived through spectral energy distribution (SED) fitting that employs a combination of galaxy and AGN templates to account for non-stellar emission (e.g., Salvato et al.~2011). For the C-COSMOS sources, we adopt the spectroscopic and photometric redshifts compiled by Civiano et al.~(2012); the former are drawn primarily come from Lilly et al.~(2009), Trump et al.~(2009), and Brusa et al.~(2010), while the latter come from the work of Salvato et al.~(2011).  The redshift and luminosity distribution of the resulting AGN sample in all three fields is shown in Figure \ref{fig-lx_z}.



\subsection{Optical High-Resolution Imaging}

To analyze the host morphologies of our AGN sample, we make use of the high resolution HST Advanced Camera for Survey (ACS) optical imaging that is publicly available in each of our three fields.  In the CDFS, we use the F850LP ($z$-band) imaging from the Great Observatories Origins Deep Survey (GOODS; Giavalisco et al.~2004), which covers the central $10^{\prime}\times16^{\prime}$ of the field.  This imaging has an exposure time of $\sim18200$ sec and reaches a limiting magnitude of m$_{AB}=28.3$ ($5\sigma$, point source, within a circular aperture of radius $0\farcs12$; Grogin et al.~2011).  In COSMOS, we use the F814W ($I_{F814W}$-band) mosaic that covers an area of roughly $77^{\prime}\times77^{\prime}$ with an exposure time of $\sim2000$ sec and which reaches a limiting magnitude of m$_{AB}=27.2$ (Koekemoer et al.~2007). 
In the EGS, we make use of the AEGIS F814W mosaic which covers a $10.1^{\prime}\times70.5^{\prime}$ region.  This imaging has an exposure time of $\sim2100$ sec and reaches a limiting magnitude of m$_{AB}=27.5$ (Davis et al.~2007).  The ACS imaging in all three fields have a pixel scale of $0\farcs03$/pixel.




\subsection{Identifying Obscured AGN}


We select obscured AGN from our parent sample using an X-ray spectral analysis that provides a measure of the line-of-sight obscuration present in each source.  The details of this spectral fitting are presented in Brightman et al.~(2014); below we briefly summarize this analysis. 

Individual source spectra were extracted using ACIS Extract (Broose et al.~2010) and lightly grouped with a minimum of one count per bin using the HEASARC tool {\tt grppha}.  The spectral fitting was carried out with {\tt XSPEC} using the Cash statistic (c-stat; Cash 1979).  
The spectral models we use are from Brightman \& Nandra (2011), which employ Monte Carlo simulations to account for Compton scattering and the geometry of the obscuring material.  They also include a self-consistent treatment of iron K$\alpha$ emission and describe spherical and torus distributions of the circumnuclear material.  

 
Four model combination are fit to each spectrum.  The first three models represent obscured emission with various torus geometries. In these models, the column density, $N_{\rm H}$, primary power-law index, $\Gamma$, and power-law normalization are free parameters.  Rather than attempting to constrain the torus opening angle from the spectra, three different cases where tested where torus opening angles were fixed at $60^{\circ}$, $30^{\circ}$ and $0^{\circ}$; here $0^{\circ}$ is essentially a $4\pi$ spherical distribution. For opening angles $>0^{\circ}$ we include a secondary power-law component, $\Gamma_{\rm scatt}$, in the fit, which represents intrinsic scattered emission, reflected by hot electrons filling the cone of the torus.  Here $\Gamma_{\rm scatt}$ is set to the primary power law index.  When the opening angle is $0^{\circ}$, we do not include this scattered component as this model represents the case where there is no escape route for the primary radiation to be scattered into the line of sight. The fourth model is a simple power-law model with two free parameters, the power-law index, $\Gamma$, and its normalization, which represents purely unobscured X-ray emission.

\begin{figure}[t]
\vspace{0.1in}
\epsscale{1.1}
\plotone{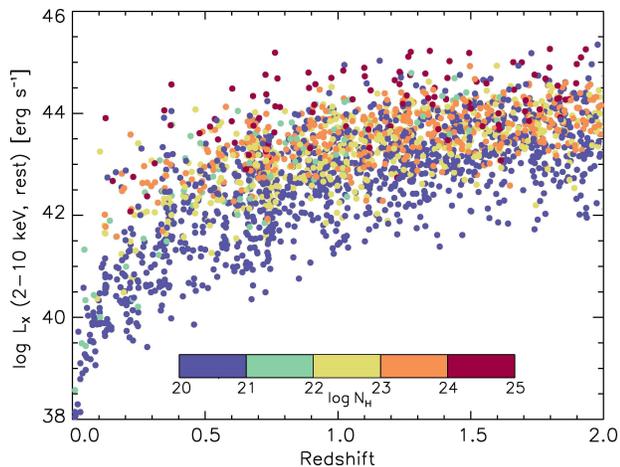}
\caption{Redshift versus luminosity for X-ray sources in the CDFS, EGS and COSMOS fields.  Sources are color coded by their level of nuclear obscuration, as determined by our X-ray spectral modeling.  X-ray luminosities are intrinsic rest-frame 2-10 keV luminosities having been corrected for absorption.
  \label{fig-lx_z_nh}}  
\end{figure}

Each of the four model combinations is fit to the source spectrum in turn with at least 100 iterations.  We adopted a critical $\Delta$c-stat of $1\times10^{-5}$ as the minimum decrease in the fit statistic required for {\tt XSPEC} to say that it has found the minimum.  The best-fitting model combination is chosen to be that which presents the lowest c-stat value after penalizing the more complex models (those with more free parameters).  However, Brightman \& Ueda (2012) have shown that for sources with less than 600 counts, large uncertainties in the spectral fits can be reduced by fixing the power-law index in the fit.  Thus, for these sources, we use a fixed value of $\Gamma=1.7$, the mean spectral index of sources with more than 600 counts.  We do, however, allow a consideration for intrinsically steep or flat spectra.  If the best-fitting model for sources with less than 600 counts, where $\Gamma$ is free, is a significantly better fit than the best-fitting model where $\Gamma$ is fixed, using the criterion of $\Delta$c-stat $>$ 2.71, we choose the model with $\Gamma$ free as the best-fitting model.  In total, the fraction of sources where $\Gamma$ is left free is 221/549, 220/937, and 232/1761 in the CDFS, EGS, and COSMOS fields, respectively.  Examples of our X-ray spectral fits in all three fields can be seen in Figure \ref{fig-spectra}.

\begin{figure*}[t]
\epsscale{1.15}
\plotone{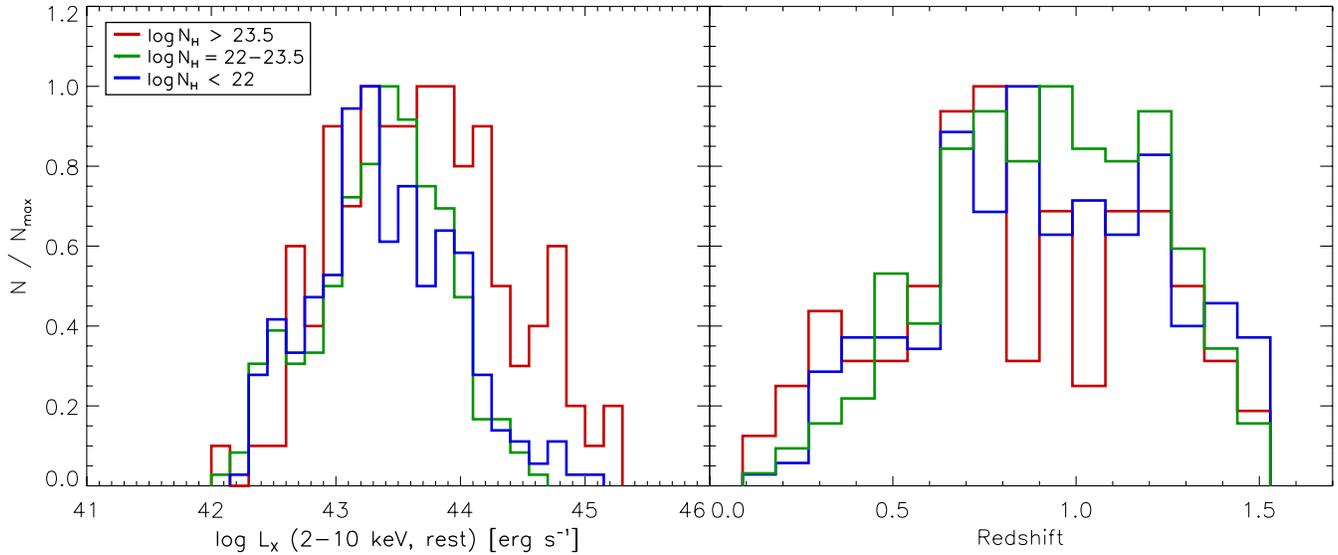}
\caption{Distribution of absorption corrected, rest-frame 2-10 keV luminosities (\emph{left}) and redshifts (\emph{right}) of the CT-AGN sample ($N_{\rm H}>10^{23.5}$ cm$^{-2}$) and our control samples of moderately obscured ($10^{22}$ cm$^{-2}$ $<N_{\rm H}<10^{23.5}$ cm$^{-2}$) and unobscured ($N_{\rm H}<10^{22}$ cm$^{-2}$) AGN. \label{fig-lx_hist}}  
\end{figure*}




 

From our spectral fits we obtain a best-fit line-of-sight column density, $N_{\rm H}$, for each source in our parent sample.  The resulting distribution of $N_{\rm H}$ versus redshift and luminosity is shown in Figure \ref{fig-lx_z_nh}.  As reported in Brightman et al.~(2014), we find that heavily obscured ($N_{\rm H} > 10^{24}$ cm$^{-2}$) sources are best fit by the torus models with opening angles of 30 or 60 degrees, i.e. not the 0 degree model, whereas sources with $10^{23}$ cm$^{-2}$ $<N_{\rm H}<10^{24}$ cm$^{-2}$ are better fit by the 0 degree model.  This is mostly due to the models being degenerate below $10^{24}$ cm$^{-2}$, so the best fit model was chosen to be the simplest one, which was the 0 degree model.  Furthermore, for Compton-thin sources, we find that the $N_{\rm H}$ values obtained using our torus models are in very good agreement with the result obtained by Lanzuisi et al.~(2013), who used simple absorption models, as would be expected for Compton-thin sources.

For our morphology study, we define a primary sample of heavily obscured AGN as those sources with $z<1.5$ and $N_{\rm H} > 10^{23.5}$ cm$^{-2}$.  We have chosen a column density limit that is lower than the canonical cutoff for Compton-thick AGN ($N_{\rm H} > 1.5\times10^{24}$ cm$^{-2}$) in order to increase our sample size of heavily obscured AGN.  In addition, our upper redshift limit is motivated by the fact that only 21\% of the heavily obscured AGN at $z>1.5$ in our three target fields have been imaged with HST/WFC3.  Without this near-infrared imaging we can not properly access the rest-frame optical morphology of galaxies beyond $z\sim1.5$.  Therefore, we limit our analysis to those sources at $z<1.5$ that fall within the EGS, GOODS and COSMOS HST/ACS imaging.  


Using these selection criteria results in a sample of 154 heavily obscured AGN at $z<1.5$ in our three target fields.  For simplicity, we will refer to these sources as our CT-AGN sample for the remainder of the paper, despite our relaxed $N_{\rm H}$ cut.  Of this sample, 21 CT-AGN are drawn from CDFS data, while 44 and 89 are detected in the EGS and COSMOS fields, respectively. 

\subsection{Control Sample Selection}



In order to compare the host morphologies of CT-AGN to that of their less obscured counterparts, we have constructed two control samples which are matched in redshift and X-ray luminosity to the CT-AGN sample, but have lower measured absorbing column densities. 
These two control samples consist of unobscured AGN with $N_{\rm H}<10^{22}$ cm$^{-2}$ and moderately obscured AGN with $10^{22}<N_{\rm H}<10^{23.5}$ cm$^{-2}$. In order to match the redshift and luminosity distributions of the samples, for each CT-AGN we randomly select two unobscured and two moderately obscured AGN that have a redshift within $\Delta z = 0.1$ and an absorption corrected X-ray luminosity within a factor of two ($0.5 \le L_{\rm X, CT} / L_{\rm X, control} \le 2$) of the CT-AGN.  For this matching we use rest-frame 2-10 keV absorption corrected luminosities, which are derived by setting $N_{\rm H}$=0 in our best-fit spectral model.  If two unique comparison AGN could not be found within this parameter range, the search range is iteratively increased by 10\%.  Because of differences in the depth of the GOODS-S, EGS and COSMOS HST/ACS imaging, the control AGN were selected separately for each region.  
%
%
The resulting luminosity and redshift distribution of the three subsamples are shown in Figure 4.  The median obscuration-corrected luminosity of the CT-AGN, moderately obscured, and unobscured subsamples are $<L_{\rm 2-10~ keV}>$ = 10$^{43.69}$, 10$^{43.40}$, and 10$^{43.34}$ erg s$^{-1}$, respectively.  

Selecting control AGN matched in luminosity is challenging because the large absorption corrections applied to the luminosities of the CT-AGN make them among the most luminous sources in our fields.  Statistically (i.e.~according to a K-S test), the luminosity distributions of the three subsamples are not perfectly matched, with the CT-AGN having a longer tail toward higher X-ray luminosities, as evidenced by their slightly higher median 2-10 keV luminosity.  However, our methodology effectively ensures that we have selected the most luminous moderately obscured and unobscured sources in each field that have similar redshifts as the CT-AGN.  Unfortunately, the only way to improve our luminosity matching would be to increase the sample size of AGN available to draw upon.  A proper redshift and luminosity matching is vital since it has been proposed that galaxy mergers play a greater role in triggering luminous AGN, while secular processes trigger lower luminosity AGN (e.g., Triester et al.~2012).  That said, we do not believe the difference in the median luminosity of the three subsamples is large enough to be the primary driver of the results presented in \S4 as we find no systematic trend between disturbed morphologies and absorption-corrected luminosity among the CT-AGN that would indicate mergers dominate the tail of the CT-AGN luminosity distribution.


Finally, since we are interested in assessing the morphology of the AGN hosts, we follow Cisternas et al.~(2011) and apply a magnitude cut of $I_{\rm F814W}< 24$ for AGN in the EGS and COSMOS fields and $z_{\rm F850LP}<24$ for sources in the CDFS.  This leaves 120, 279, 282 sources in the Compton-thick, moderately obscured, and unobscured AGN samples, respectively\footnote{Our $N_{\rm H}$ cut of $10^{23.5}$ cm$^{-2}$ was chosen to ensure that roughly half (61/120) of our final CT-AGN sample have $N_{\rm H} > 10^{24}$ cm$^{-2}$, and are therefore truly Compton-thick.  It should be noted, though, that given the large uncertainties on our $N_{\rm H}$ estimates, even sources with $N_{\rm H}\sim10^{23.5}$ cm$^{-2}$ could still be consistent with being Compton-thick.}.  


\begin{figure*}[t]
\epsscale{1.0}
\plotone{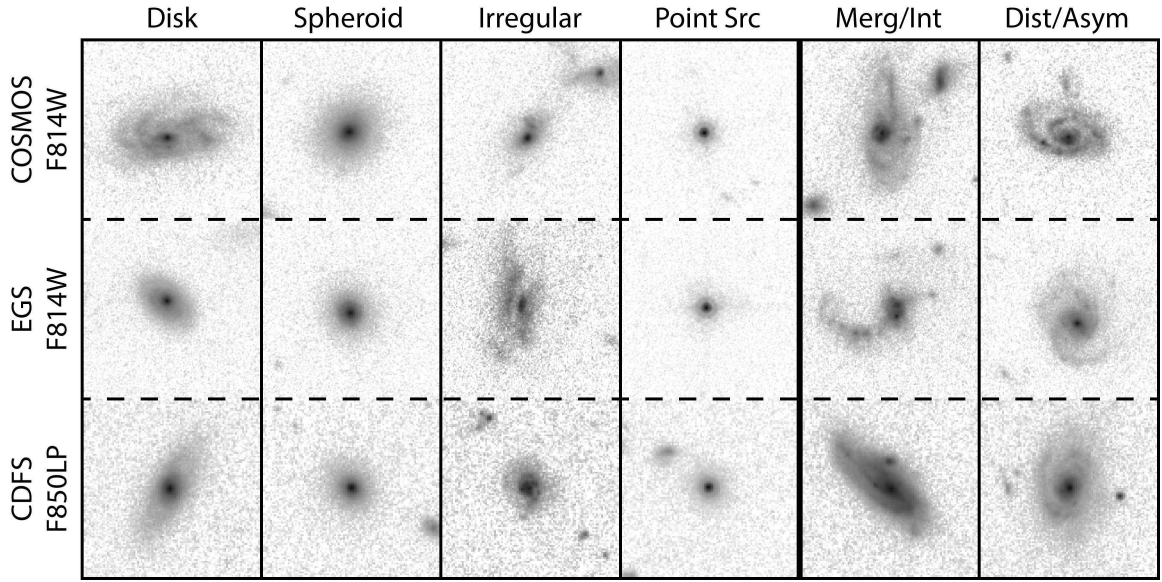}
\caption{Examples of AGN host galaxies in each morphology and disturbance class of our visual classification scheme.  While the \emph{Disk}, \emph{Spheroid}, \emph{Irregular}, \emph{Point-like} classifications are mutually exclusive, the \emph{Disturbed/Asymmetric} class is a superset of the \emph{Merger/Interaction} class, as it includes train-wreck mergers and galaxies that exhibit only minor disturbances.  See \S3 for details.
  \label{MorphScheme}}  
\end{figure*}

\section{Morphology Classification} 

Host morphologies of the CT-AGN and control AGN were assessed through visual inspection using a classification scheme similar to the one presented in Kocevski et al.~(2012) and Kartaltepe et al.~(2015).  These inspections were carried out by the lead author, D.K., and performed blind using the reddest HST/ACS bands available in each field, namely the F814W band in the EGS and COSMOS fields and the F850LP band in the CDFS.  In addition, F606W imaging was used to provide supplemental color information for sources in the EGS and CDFS (similar imaging is not available in the COSMOS field).  The size of each thumbnail image was set to cover roughly 100 kpc on a side at the redshift of each AGN and ranged from $12^{\prime\prime}-16^{\prime\prime}$.
Because of differences in the bands used in each field and the depth of the available imaging, control AGN were drawn from the same field as their matched CT-AGN and the subsequent classifications were carried out separately for each field.\footnote{Where possible, we have compared the classification of D.K. to those of the CANDELS collaboration (where each galaxy was inspected by an average of 4 unique classifiers; Kartaltepe et al.~2015), and we find excellent ($>90\%$) agreement between the two.}

For each AGN host, we classified the morphological type of the galaxy and the degree to which it is disturbed.  The possible morphologies were: \emph{Disk}, \emph{Spheroid}, \emph{Irregular/Peculiar}, \emph{Point-like}. These classes are mutually exclusive and only the predominate morphology of each galaxy was noted.  For example, disk galaxies with a substantial bulge component would simply be classified as disks in this system.  This differs from the scheme used in Kocevski et al.~(2012), where bulge and disk dominated late-type galaxies were differentiated. The change was made to mitigate the effects of moderate AGN contamination, which can mimic an increase in the bulge-to-disk ratio of a galaxy.  In this scheme, as long as an extended disk is visible, regardless of the level of nuclear AGN contamination, the galaxy is classified as a \emph{Disk}.  This is physically motivated by the fact that disks are easily destroyed in major mergers and take a considerable amount of time to reform (Robertson et al.~2006).  Therefore the presence of a disk constrains, to a certain extent, the past merger history of a given galaxy\footnote{It is important to note that while disks can reform following a merger under the right circumstances (when they are sufficiently gas rich and have favorable initial orbital parameters), disk survival is most efficient at low galaxy masses and generally require conditions that suppress strong inflows toward the galaxy center (see Springel \& Hernquist 2005; Robertson et al. 2006; Hopkins et al. 2009).  This effectively prevents strong bulge growth and AGN fueling; the opposite of the regime we are interested in here.}.

To gauge the degree to which a galaxy is disturbed, three disturbance classifications were used:

\vspace{0.1in}
\hangindent=0.25in \hangafter=1 $\bullet$ \emph{Merger/Interaction:} Two distinct galaxies showing interaction features such as tidal arms or a single train-wreck system exhibiting strong distortions.

\hangindent=0.25in \hangafter=1 $\bullet$ \emph{Disturbed/Asymmetric:} All galaxies in the \emph{Merger/ Interaction} class plus single asymmetric or disturbed galaxies with no visible interacting companion. 

\hangindent=0.25in \hangafter=1 $\bullet$ \emph{Undisturbed:} None of the above. \\


\begin{figure*}[t]
\epsscale{1.15}
\plotone{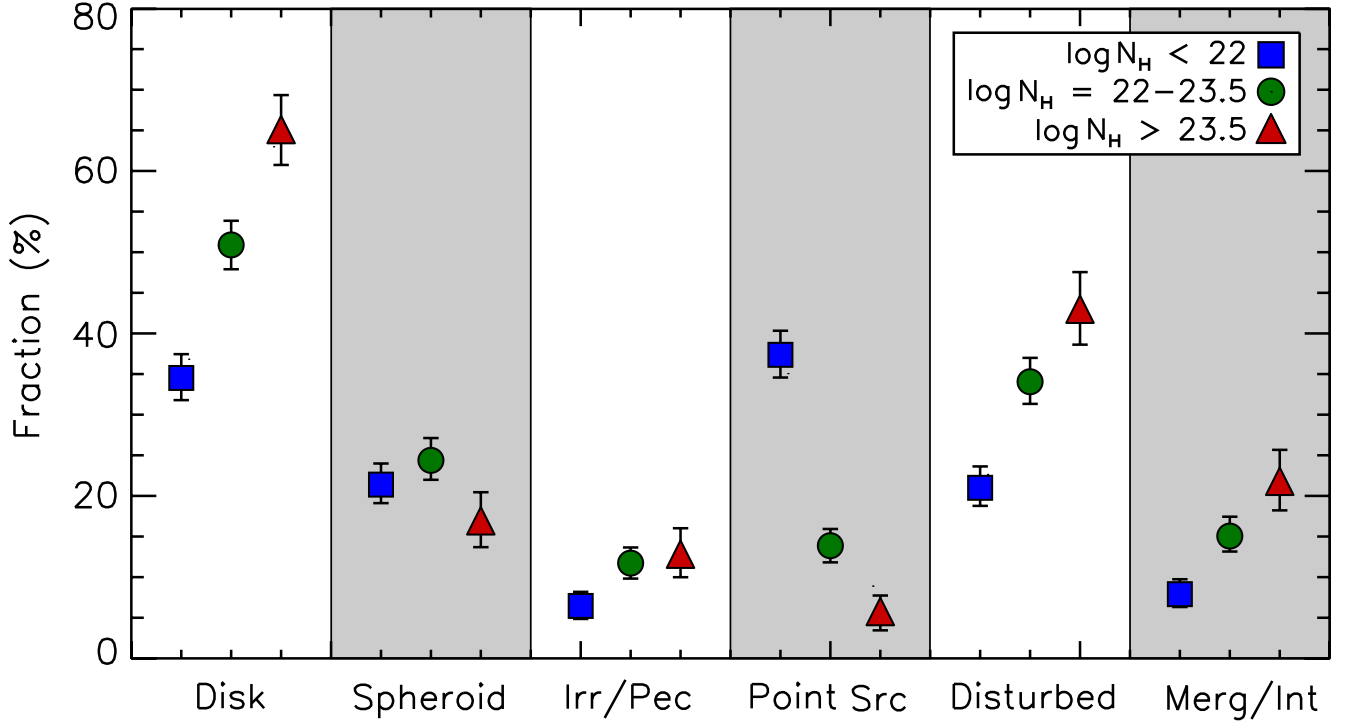}
\caption{Fraction of AGN hosts at $0.5<z<1.5$ assigned to various morphology and disturbance classes as a function of their nuclear obscuration.  We find that the hosts of heavily obscured AGN are more likely to be disks and have disturbed morphologies relative to the hosts of unobscured AGN with the same redshift and absorption-corrected X-ray luminosity.
  \label{fig-summary-wPS}}  
\end{figure*}

In this scheme the \emph{Merger/Interaction} class includes train-wreck mergers that have multiple nuclei and/or strong distortions in a single coalescing system, as well as disturbed galaxies with an interacting companion.  The \emph{Disturbed/Asymmetric} class, however, serves as a more liberal selection of galaxies that may have experienced an interaction in the recent past.  This class includes any galaxy which has a distorted or asymmetric light profile, even those with no visible interacting companion.  As a result, these classes are not mutually exclusive.  These classes are similar to the \emph{Disturbed I} and \emph{Disturbed II} classes used in Kocevski et al.~(2012), respectively.  Examples of AGN host galaxies in each of our morphology classes can be seen in Figure \ref{MorphScheme}.

It should be noted that unresolved AGN hosts (those classified as having \emph{Point-like} morphologies) are by definition classified as \emph{Undisturbed} in this system.  As a result, any AGN subsample that has a high \emph{Point-like} fraction will also have a high \emph{Undisturbed} fraction.  Alternatively, one could argue that the disturbance level of \emph{Point-like} sources is not measurable and should not be classified as \emph{Undisturbed}.  In the following section, we present our results using both approaches: first including \emph{Point-like}/\emph{Undisturbed} sources in our analysis and then excluding them completely.  While our primary results do not change, the statistical significance of our findings do change as a result of having fewer AGN in our final sample.




\section{Results}

The fraction of AGN hosts in each of our morphology classes versus their level of nuclear obscuration is shown in Figure \ref{fig-summary-wPS} and listed in Table 1.  The error bars on each fraction reflect the 68.3\% binomial confidence limits given the number of sources in each category, calculated using the method of Cameron et al.~(2010).  For our sample of CT-AGN we find that $65.3^{+4.1}_{-4.6}\%$ have predominately disk-like morphologies.  This includes disks with and without a central bulge.  A smaller fraction, $16.5^{+3.9}_{-2.8}\%$, are classified as spheroidal, whereas $12.4^{+3.6}_{-2.4}\%$ are found to have peculiar or irregular morphologies such that neither a prominent disk or spheroidal component could be discerned.  Only a small fraction, $5.0^{+2.8}_{-1.3}\%$, of the CT-AGN are classified as point-like.  This may be expected if heavy nuclear obscuration is blocking emission from the central engine in these sources.

For our control sample of unobscured AGN ($N_{\rm H}<10^{22}$ cm$^{-2}$) we find a lower disk fraction ($34.5^{+2.9}_{-2.7}\%$) relative to the CT-AGN hosts, a slightly higher spheroid fraction ($21.4^{+2.6}_{-2.2}\%$), and a lower irregular fraction ($6.4^{+1.8}_{-1.2}\%$).  Unlike their heavily obscured counterparts, a much larger fraction of the unobscured sources appear point-like in the ACS imaging, accounting for $37.4^{+3.0}_{-2.8}\%$ of the host morphologies.  The hosts of the moderately obscured ($10^{22}$ cm$^{-2}$ $<N_{\rm H}<10^{23.5}$ cm$^{-2}$) control sample have morphologies that lie between the two extremes of the Compton-thick and unobscured AGN.  Here disks make up $50.9^{+3.0}_{-3.0}\%$ of the population, spheroids $24.4^{+2.7}_{-2.4}\%$ and irregulars $11.5^{+2.2}_{-1.6}\%$.  We find an increased point-like fraction ($13.6^{+2.3}_{-1.8}\%$) relative the CT-AGN population, however this fraction is lower than that found in the unobscured control sample. 

\begin{figure*}[t]
\epsscale{1.15}
\plotone{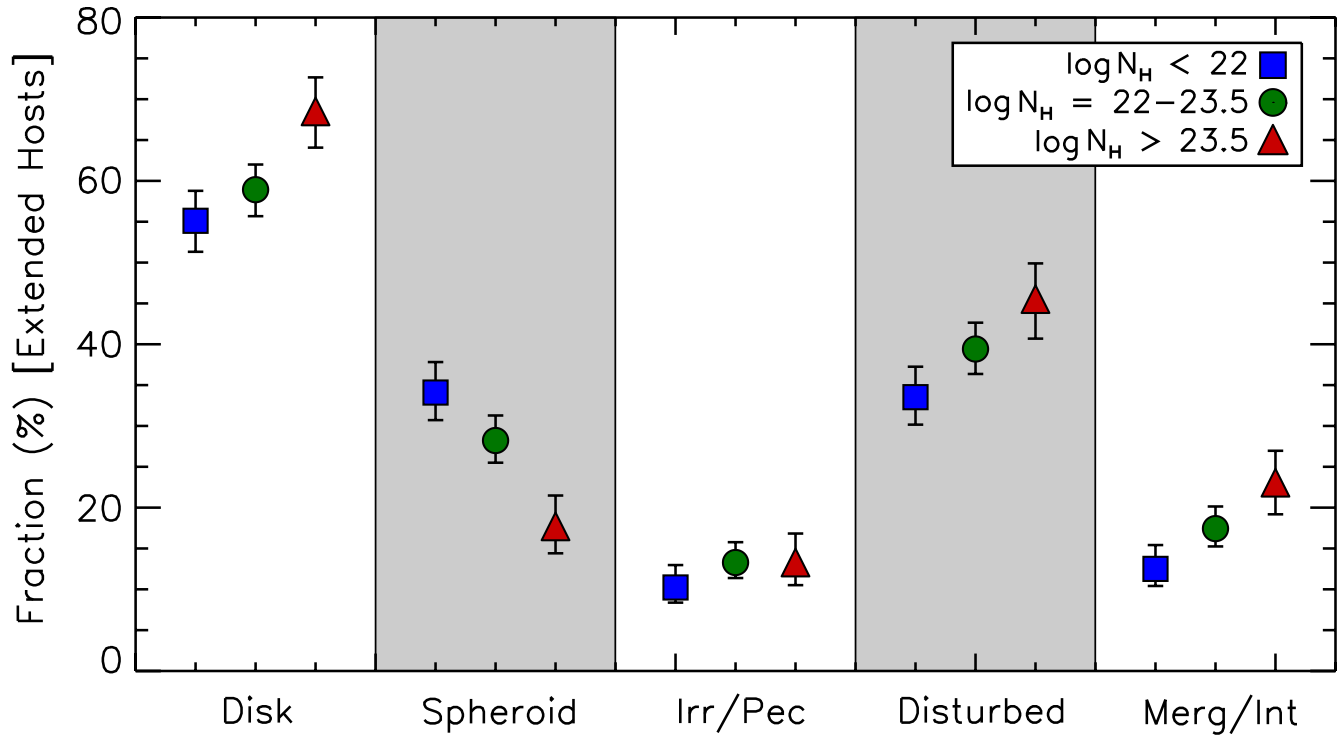}
\caption{Fraction of extended AGN hosts at $0.5<z<1.5$ assigned to various morphology and disturbance classes as a function of their nuclear obscuration.  This plot is the same as Figure \ref{fig-summary-noPS}, however point sources are now excluded from the analysis.  We again find that the hosts of heavily obscured AGN are more likely to be disks and show some morphological disturbance relative to our unobscured control sample, albeit at a reduced statistical significance.  See \S4 for details.
  \label{fig-summary-noPS}}  
\end{figure*}

The fraction of AGN with disturbed morphologies in each of our three subsamples is also shown on the right side of Figure \ref{fig-summary-wPS}.  We find a statistically significant increase in the \emph{Merger/Interaction} fraction versus AGN obscuration, rising from $7.8^{+1.9}_{-1.3}\%$ among the unobscured AGN to $15.1^{+2.4}_{-1.9}\%$ for the moderately obscured AGN and $21.5^{+4.2}_{-3.3}\%$ for the CT-AGN sample.  The increase in the merger fraction of the CT-AGN relative to their unobscured counterparts is significant at the 3.8$\sigma$ level.  If we include any galaxy that has a distorted or asymmetric light profile, the overall disturbed fraction increases in all three samples, but the trend with obscuration remains.  The \emph{Disturbed/Asymmetric} fraction increases from $21.0^{+2.6}_{-2.2}\%$ among the unobscured AGN to $34.1^{+2.9}_{-2.7}\%$ for the moderately obscured AGN and $43.0^{+4.6}_{-4.4}\%$ for the CT-AGN sample.  Here the difference in the disturbed fraction of the CT-AGN relative to the unobscured AGN is significant at the 4.4$\sigma$ level.

As discussed in \S3, the high point source fraction among the unobscured AGN may artificially drive the disturbed fraction down for that subsample since unresolved hosts are classified as \emph{Undisturbed} by default.  To account for this, we have excluded all unresolved hosts from our analysis and present the resulting morphology and disturbance fractions in Figure \ref{fig-summary-noPS}.  When considering only extended hosts, we find that the disk fraction of the three subsamples is in much greater agreement, although the CT-AGN are still found in disk hosts more often than the unobscured AGN.  We find the disk fraction steadily increases from $55.1^{+3.6}_{-3.8}\%$ among the unobscured AGN to $58.9^{+3.1}_{-3.2}\%$ for the moderately obscured AGN and $68.7^{+4.0}_{-4.6}\%$ for the CT-AGN sample.  This trend reverses for the spheroid fraction, which steadily decreases with obscuration.  Here the spheroid fraction decreases from $34.1^{+3.7}_{-3.4}\%$ among the unobscured AGN to $28.2^{+3.0}_{-2.7}\%$ for the moderately obscured AGN and $17.4^{+4.1}_{-3.0}\%$ for the CT-AGN sample.  The irregular fraction is low for all three subsamples and consistent with showing no trend with obscuration.  

\begin{figure}[t]
\epsscale{1.15}
\plotone{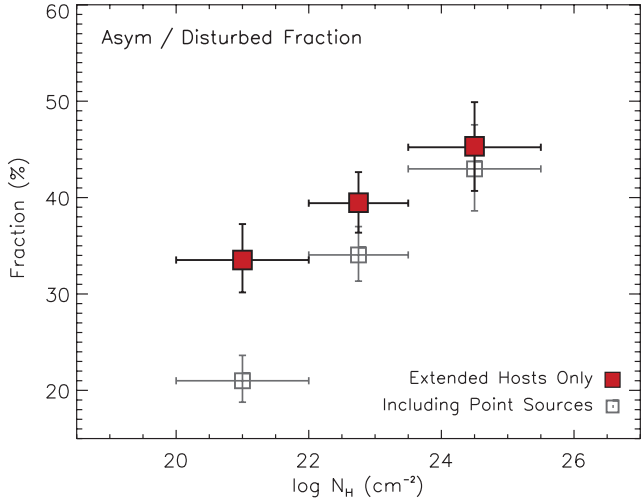}
\caption{Fraction of AGN hosts in the \emph{Disturbed/Asymmetric} class as a function of nuclear obscuration, with point sources included (\emph{open box}) and excluded (\emph{filled box}) from the analysis.  In both cases, the hosts of heavily obscured AGN are more likely to be classified as disturbed relative to unobscured AGN with similar redshifts and luminosities.  However, when point sources are excluded, the statistical significance of this difference drops from 4.4$\sigma$ to 2.3$\sigma$. \label{fig-dist}}  
\end{figure}

\begin{figure}[t]
\epsscale{1.15}
\plotone{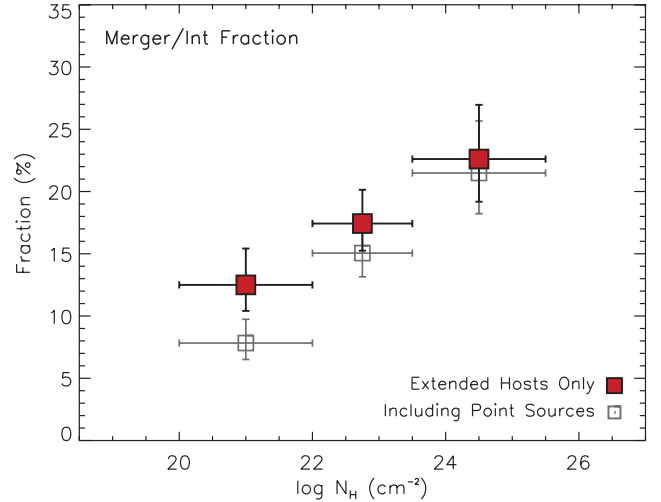}
\caption{Fraction of AGN hosts in the \emph{Merger/Interaction} class as a function of nuclear obscuration, with point sources included (\emph{open box}) and excluded (\emph{filled box}) from the analysis. In both cases, the hosts of heavily obscured AGN are more likely to be classified as being involved in a merger or interaction relative to unobscured AGN with similar redshifts and luminosities.  However, when point sources are excluded, the statistical significance of this difference drops from 3.8$\sigma$ to 2.5$\sigma$.
  \label{fig-merg}}  
\end{figure}

The fraction of AGN in extended hosts that exhibit a morphological disturbance is shown in Figures \ref{fig-dist} and \ref{fig-merg}.  The correlation between merger fraction and obscuration is still evident when excluding point sources, however the statistical significance of the increase drops from 3.8$\sigma$ to 2.5$\sigma$.  The \emph{Merger/Interaction} fraction is now $12.5^{+2.9}_{-2.1}\%$, $17.4^{+2.7}_{-2.2}\%$, and $22.6^{+4.3}_{-3.4}\%$, for the unobscured, moderately obscured, and CT-AGN samples.  A similar trend is found for the \emph{Disturbed/Asymmetric} fraction, which increases from $33.5^{+3.7}_{-3.4}\%$ among the unobscured AGN to $39.4^{+3.2}_{-3.1}\%$ for the moderately obscured AGN and $45.2^{+4.7}_{-4.5}\%$ for the CT-AGN sample.  Here the statistical significance of the increase is now 2.3$\sigma$.

In summary, we find an increasing disk fraction and a decreasing spheroid fraction with increasing nuclear obscuration among AGN at $0.5<z<1.5$.  In addition, we find that the fraction of AGN with disturbed host morphologies increases as a function of obscuration.  This increase is found whether we consider only train-wreck mergers and galaxies with clear interacting companions or any galaxy showing an asymmetric light profile.  It is also present regardless of whether we exclude unresolved host galaxies from our analysis, albeit at a reduced statistical significance.

\section{Discussion}

Using a sample of heavily obscured AGN identified by their X-ray spectral signatures, we find a correlation between disturbed host morphology and nuclear obscuration at fixed AGN luminosity and redshift.  In this section we discuss the implications of this result in terms of both the AGN unification model (\S5.1) and the role that mergers play in fueling SMBH growth (\S5.2).  In addition, we conclude the section with a discussion of several important caveats to keep in mind when interpreting our findings (\S5.3).

\subsection{Implications for the AGN Unification Model}

The standard unification paradigm invokes a torus-like structure that obscures the central engine for some sight lines and not for others, producing the two observed AGN types. In this scheme AGN obscuration is largely dependent on the viewing angle of the observer (Antonucci 1993; Urry \& Padovani 1995; Tran 2003) and therefore all AGN would sample the same parent population of host galaxies, regardless of their level of obscuration.  In other words, there should be no correlation between heavy nuclear obscuration and disturbed host morphologies.  Alternatively, obscured SMBH growth may be a distinct phase in the co-evolution of AGN and their hosts, specifically one in which the central engine goes through rapid growth phase following a merger event (Sander et al.~1998; Hopkins et al.~2005, 2008).  This is supported by the findings of Draper \& Ballantyne 2010, who suggest that a vast majority of AGN accreting near the Eddington limit must be hidden by Compton-thick obscuration based on the observed space density of CT-AGN and their contribution to the CXB.   Furthermore, hydrodynamical merger simulations predict that this obscured phase should coincide with the most morphologically disturbed phase of a galaxy interaction (Cattaneo et al.~2005; Hopkins et al.~2008).  Therefore, merger-driven co-evolution models predict that there should be a strong dependence between obscuration and host properties such as morphology.  

\begin{figure*}[t]
\epsscale{1.15}
\plotone{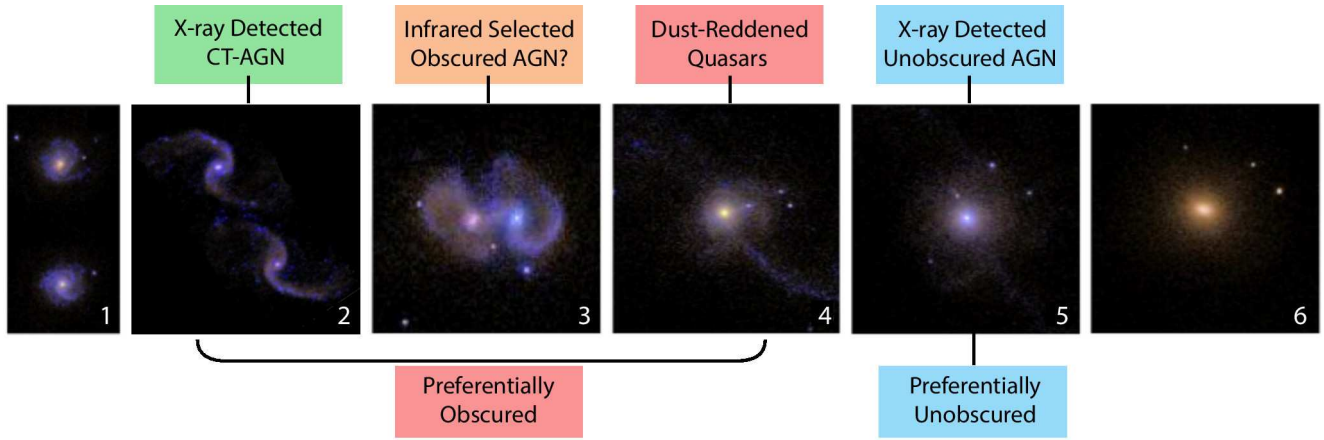}
\caption{AGN fueling models have proposed that obscured SMBH growth is a distinct phase in an evolutionary sequence following a merger event.  Given that our CT-AGN are largely hosted by disturbed, disk-dominated galaxies, we propose that we are catching these systems near the start of this evolutionary sequence, before the disk structure of these galaxies is substantially disturbed or destroyed.  The previously proposed location of obscured quasars and infrared-selected AGN along this sequence is also shown. See \S5.1 for details.  \label{fig-lotz}}  
\end{figure*}

The fact that we observe a different disk/spheroid and merger fraction versus obscuration indicates that viewing angle cannot be the only thing differentiating our three AGN samples, as the unification model would suggest.  That is not to say that viewing angle plays no part in obscuring the CT-AGN in our sample, only that interactions play a greater role in fueling their activity relative to the unobscured AGN in our control samples.  This finding appears to support an evolutionary scenario, in which an increased fraction of the CT-AGN are heavily obscured as a result of a growth phase triggered by a galaxy interaction in the recent past.  Given that the CT-AGN are hosted by largely disk-dominated galaxies, we propose that we are catching these systems near the start of this evolutionary sequence, before the disk structure of these galaxies is substantially disturbed or destroyed.  This may be due to increasing obscuration levels as the merger sequence progresses.  For example, if the covering fraction of the obscuring torus is higher for sources further along in the merger sequence, then the fraction of X-ray photons scattered into our line-of-sight would decrease and we would therefore not be sensitive to the most disturbed sources.  Our proposed location for the CT-AGN sample in a possible evolutionary sequence is illustrated in Figure \ref{fig-lotz}.

Previous studies have reached similar conclusions regarding the transitional nature of obscured AGN.  For example, the hosts of dust-reddened quasars (Urrutia et al. 2009; Glikman et al. 2004, 2012) show a high incidence of merger activity and a disturbance fraction that increases with increasing obscuration (Urrutia et al. 2008; Glikman et al.~2015). These quasars are intrinsically more luminous than the CT-AGN in our sample and are preferentially found in spheroid-dominated hosts.  It is therefore thought that these quasars are detected in the final stages of emerging from their dusty cocoons and near the end of the evolutionary sequence outlined in Figure \ref{fig-lotz} (Urrutia et al.~2012, Glikman et al.~2012, Banerji et al.~2012).

Our results also agree with the recent findings of Donley et al.~(2015) and Juneau et al.~(2013), who examined the morphologies and star formation activity, respectively, of obscured AGN selected by their mid-infrared colors and emission line properties.  In the former, the hosts of IRAC power-law selected AGN (Donley et al.~2008) are found to be more disturbed than their X-ray selected, and presumably less obscured, counterparts.  In the latter, the obscured AGN fraction is found to be higher among galaxies with elevated specific star formation rates, which the authors argue may be due to recent galaxy interactions.



On the other hand, our findings are at odds with the results of Schawinski et al.~(2012).  Here the authors examined the morphology of DOGs in the Extended CDFS, which are thought to host heavily obscured quasars based on the X-ray stacking analysis of Treister et al.~(2009).  This study found that only a small fraction ($\sim4\%$) of DOGs at $1<z<3$ show signs of recent merger activity.  It is worth noting, however, that only one of our CT-AGN in the CDFS would be selected as an obscured AGN via the infrared excess method employed by Schawinski et al.~(2012).  This is consistent with the findings of Comastri et al.~(2011), who noted that X-ray detected CT-AGN are not readily picked up by standard mid-infrared selection techniques.  We therefore suspect our conflicting results are due to substantial differences in our parent samples and we caution against direct comparisons of the two studies.

\subsection{Implications for the AGN-Merger Connection}

Galaxy mergers have long been proposed as a possible triggering mechanism for AGN activity, however there is a growing consensus that most moderate-luminosity, X-ray selected AGN show no signs of recent merger activity based on their host morphologies (Grogin et al.~2005, Cisternas et al.~2011, Schawinski et al.~2011, Kocevski et al.~2012, Villforth et al.~2014).  In fact, recent results from the CANDELS survey suggest that stochastic fueling by secular processes or disk instabilities play a greater role in fueling SMBH growth at $z>1$ than previously predicted by AGN fueling models (Kocevski et al.~2012).  This is likely due to the increasing gas fraction of galaxies at high redshifts (e.g.~Tacconi et al.~2010), which acts to increase the duty cycle of distant, stochastically-fed AGN (Hopkins et al.~2007, Johansson et al.~2009).


However, even with these new observational constraints, AGN fueling models continue to predict that the integrated total SMBH growth in the Universe should be dominated by merger-induced fueling.  For example, the semi-emprical fueling model of Hopkins, Kocevski, \& Bundy (2014), which incorporates both stochastic and merger-induced fueling modes, finds that while non-merger processes may dominate the AGN population by numbers, only $\sim30\%$ of the total AGN luminosity density and SMBH mass density is the result of stochastic fueling.  The predicted contribution is strongly mass and luminosity-dependent, with mergers playing an increasingly important role in fueling high-mass ($M_{\rm BH}>10^{7} M_{\odot}$) SMBH growth and high luminosity ($L_{\rm bol}>10^{12} L_{\odot}$) AGN.  A similar conclusion was reached by Draper \& Ballantyne (2012) using AGN population synthesis modeling to determine the importance of different AGN triggering mechanisms.  

It is conceivable that some of the merger-induced fueling that is predicted may have been missed by past studies of the \emph{Chandra} deep fields, given the few high-luminosity AGN present in these fields (e.g.~Triester et al.~2012).  However, our results indicate that a portion of this merger-triggered activity may also be hidden among heavily obscured AGN.  This implies that past studies may have missed the AGN population where the AGN-merger connection is expected to be the strongest.  Whether there is sufficient merger-fueled SMBH growth occurring among heavily obscured and high luminosity AGN to match what is predicted by the Hopkins et al.~(2014) fueling model remains to be determined.  A key to testing this will be identifying additional CT-AGN at $z\sim1-2$ in order to determine what fraction of this obscured growth remains undetected.



\subsection{Caveats \& Future Work}

Having discussed the possible implications of our results, it is worth keeping in mind a couple of important caveats.  First, the difference in the disturbed fraction between the obscured and unobscured AGN is statistically significant only when point-like hosts are included in our analysis.  When we consider only hosts with extended morphologies, the elevated merger fraction among the CT-AGN is significant at only the $2.5\sigma$ level.  A larger sample of distant, obscured AGN will need to be identified in order to confirm our findings with greater statistical confidence.  This will soon be possible as a result of the X-UDS Chandra Legacy Survey (co-PIs G.~Hasinger and D.~Kocevski), which is obtaining deep (1.25 Msec) and wide ($22^{\prime}\times22^{\prime}$) X-ray observations of the UKIRT Infrared Deep Sky Survey (UKIDSS) Ultra-deep Survey field (UDS; Lawrence et al.~2007; Cirasuolo et al.~2007).  This dataset, when combined with the existing ACS and WFC3 imaging from CANDELS, will substantially increase the number of CT-AGN at $z\sim1$ available for study.  It will also increase the number of CT-AGN at $z\sim2$ that have rest-frame optical imaging, allowing us to extend our study to this redshift for the first time.



The second caveat relates to a possible connection between the elevated disk and disturbed fractions among the CT-AGN.  If morphological disturbances are easier to visually detect in late-type systems, then it is possible that the increased \emph{Disturbed/Asymmetric} fraction among the CT-AGN is simply a reflection of their higher \emph{Disk} fraction relative to the control samples.  We do not believe this is the case, as the CT-AGN also show an increased \emph{Merger/Interaction} fraction, a classification that requires a visible interacting neighbor or a train-wreck morphology.  In other words, highly disruptive events that should be detectable in early-type galaxies as well as their late-type counterparts.  Nonetheless, further work is needed to determine what fraction of interactions may be missed among spheroidal hosts.  The CANDELS team is actively pursuing this work by visually classifying simulated interactions using a classification scheme similar to the one used in this study.


\begin{center}
\tabletypesize{\scriptsize}
\begin{deluxetable*}{llcc}
\tablewidth{0pt}
\tablecaption{Visual Classification Results}
\tablecolumns{4}
\tablehead{\colhead{}               & \colhead{Unobscured AGN} & \colhead{Moderately-Obscured AGN} & \colhead{Compton-Thick AGN}  \\  
           \colhead{} & \colhead{($N_{\rm H}<10^{22}$ cm$^{-2}$)}        & \colhead{($10^{22}<N_{\rm H}<10^{23.5}$ cm$^{-2}$)}  & \colhead{($N_{\rm H}>10^{23.5}$ cm$^{-2}$)} \\
		   \colhead{Classification} & \colhead{All Hosts/Extended Hosts} & \colhead{All Hosts/Extended Hosts} & \colhead{All Hosts/Extended Hosts}  }
		   \startdata
		   
Disk                & $34.5^{+2.9}_{-2.7}\%$ \hspace{0.08in} $55.1^{+3.6}_{-3.8}\%$  &  $50.9^{+3.0}_{-3.0}\%$ \hspace{0.08in} $58.9^{+3.1}_{-3.2}\%$  &  $65.3^{+4.1}_{-4.6}\%$ \hspace{0.08in} $68.7^{+4.0}_{-4.6}\%$     \\
Spheroid            & $21.4^{+2.6}_{-2.2}\%$ \hspace{0.08in} $34.1^{+3.7}_{-3.4}\%$  &  $24.4^{+2.7}_{-2.4}\%$ \hspace{0.08in} $28.2^{+3.0}_{-2.7}\%$  &  $16.5^{+3.9}_{-2.8}\%$ \hspace{0.08in} $17.4^{+4.1}_{-3.0}\%$     \\ 
Irregular           & $06.4^{+1.8}_{-1.2}\%$ \hspace{0.08in} $10.2^{+2.7}_{-1.9}\%$  &  $11.5^{+2.2}_{-1.6}\%$ \hspace{0.08in} $13.3^{+2.5}_{-1.9}\%$  &  $12.4^{+3.6}_{-2.4}\%$ \hspace{0.08in} $13.0^{+3.8}_{-2.5}\%$     \\ 
Point-like          & $37.4^{+3.0}_{-2.8}\%$ \hspace{0.08in} ------------            &  $13.6^{+2.3}_{-1.8}\%$ \hspace{0.08in} ------------            &  $05.0^{+2.8}_{-1.3}\%$ \hspace{0.08in} ------------               \\ 
Disturbed/Asym      & $21.0^{+2.6}_{-2.2}\%$ \hspace{0.08in} $33.5^{+3.7}_{-3.4}\%$  &  $34.1^{+2.9}_{-2.7}\%$ \hspace{0.08in} $39.4^{+3.2}_{-3.1}\%$  &  $43.0^{+4.6}_{-4.4}\%$ \hspace{0.08in} $45.2^{+4.7}_{-4.5}\%$     \\  
Merger/Interaction  & $07.8^{+1.9}_{-1.3}\%$ \hspace{0.08in} $12.5^{+2.9}_{-2.1}\%$  &  $15.1^{+2.4}_{-1.9}\%$ \hspace{0.08in} $17.4^{+2.7}_{-2.2}\%$  &  $21.5^{+4.2}_{-3.3}\%$ \hspace{0.08in} $22.6^{+4.3}_{-3.4}\%$     \\  

\vspace*{-0.075in}
\enddata
\end{deluxetable*}
\end{center}

\section{Conclusions}


We have used HST/ACS imaging to examine the morphologies of galaxies hosting heavily obscured AGN at $z\sim1$ in order to test whether obscured SMBH growth at this epoch is linked to major merger events.  Using the X-ray spectral analysis of Brightman et al.~(2014), we select 154 heavily obscured AGN with $N_{\rm H}>10^{23.5}$ cm$^{-2}$ and $z<1.5$ in the CDFS, EGS, and COSMOS fields.  To determine if these AGN are triggered by galaxy interactions more often than less obscured AGN, we construct two control samples composed of moderately obscured ($10^{22}<N_{\rm H}<10^{23.5}$) and unobscured ($N_{\rm H}<10^{23.5}$) AGN.  These samples are matched in redshift and intrinsic X-ray luminosity to the heavily obscured AGN sample.  To determine the morphology of the host galaxies, we employ a visual classification scheme similar to the one used in Kocevski et al.~(2012) and by the CANDELS collaboration.  We assess both the predominant morphology of each host galaxy and the level of disturbance that is visible.  Based on our visual classifications, we find:

\begin{enumerate}
  \item The heavily obscured AGN are predominantly hosted by late-type galaxies; $65.3^{+4.1}_{-4.6}\%$ are classified as \emph{Disks}, while only $16.5^{+3.9}_{-2.8}\%$ are classified as \emph{Spheroids}.  This disk fraction is elevated relative to our control samples of moderately obscured and unobscured AGN, which have disk fractions of $50.9^{+3.0}_{-3.0}\%$ and $34.5^{+2.9}_{-2.7}\%$, respectively.  All three samples have a low \emph{Irregular/Peculiar} fraction, which ranges from $6.4^{+1.8}_{-1.2}\%$ for the unobscured AGN to $16.5^{+3.9}_{-2.8}\%$ for the most heavily obscured.  
  \item We find a statistically significant increase in the fraction of disturbed hosts versus AGN obscuration.  Roughly  $21.5^{+4.2}_{-3.3}\%$ of the Compton-thick AGN have highly disturbed host morphologies and fall in the \emph{Merger/Interaction} class. This is true for only $7.8^{+1.9}_{-1.3}\%$ of the unobscured AGN; a difference that is significant at the 3.8$\sigma$ level.  This trend with obscuration remains when we include galaxies that exhibit any minor disturbance or asymmetry in their morphology.  Here the  \emph{Disturbed/Asymmetric} fraction increases from $21.0^{+2.6}_{-2.2}\%$ for the unobscured AGN to $34.1^{+2.9}_{-2.7}\%$ for the moderately obscured AGN and $43.0^{+4.6}_{-4.4}\%$ for the Compton-thick sample.  The statistically significance of this increase is 4.4$\sigma$.
  \item We find that the incidence of \emph{Point-like} morphologies is inversely proportional to obscuration, as might be expected if heavy nuclear obscuration is blocking emission from the central engine.  To account for any biases this may introduce, we excluded all unresolved hosts from our samples and repeated our analysis.  When considering only extended hosts, we find that the disk fraction of the three subsamples is in much better agreement, although the heavily obscured AGN are still found in disk hosts more often than their unobscured counterparts ($68.7^{+4.0}_{-4.6}\%$ versus $55.1^{+3.6}_{-3.8}\%$).  Furthermore, the correlation between merger fraction and obscuration is still evident when excluding point sources, however at a reduced statistical significance.  The \emph{Merger/Interaction} fraction increases from $12.5^{+2.9}_{-2.1}\%$ to $22.6^{+4.3}_{-3.4}\%$ for the unobscured and heavily obscured samples, respectively; a difference that is now significant at the 2.5$\sigma$ level.  A similar trend is found for the \emph{Disturbed/Asymmetric} fraction, which increases from $33.5^{+3.7}_{-3.4}\%$ among the unobscured AGN to $45.2^{+4.7}_{-4.5}\%$ for the Compton-thick sample.  Here the statistical significance is 2.3$\sigma$. 
\end{enumerate}

The fact that we observe a different disk/spheroid fraction versus obscuration indicates that viewing angle cannot be the only thing differentiating our three AGN samples, as a simple unification model would suggest.  The increased fraction of disturbed morphologies with obscuration would appear to support an evolutionary scenario, in which Compton-thick AGN are a distinct phase where the central SMBH undergoes rapid, obscured growth following a merger/interaction event.  Given that our heavily obscured AGN are hosted by disk-dominated galaxies, we propose that we are catching these systems near the start of this evolutionary sequence, before their disk structure is destroyed.  Our findings also suggest that some of the merger-triggered SMBH growth that is predicted by AGN fueling models may be hidden among heavily obscured, Compton-thick AGN, as previous studies of dust-reddened quasars have proposed.  That said, a larger sample of distant, obscured AGN will need to be studied in order to confirm our findings with greater statistical confidence, especially among extended AGN hosts.  This will soon be possible as a result of the UDS Chandra Legacy Survey, which will allow us to extend this work to $z\sim2$.

\vspace{0.5in}
Support for Program number HST-GO-12060 was provided by NASA through a grant from the Space Telescope Science Institute, which is operated by the Association of Universities for Research in Astronomy, Incorporated, under NASA contract NAS5-26555.

\bibliography{}

\end{document}